\documentclass[aps,amsfonts,prl,twocolumn,showpacs]{revtex4-1}
\usepackage{epsfig,amsmath,amssymb,bm,epsf,graphics,psfrag,verbatim,subfigure}

\def\beq{\begin{equation}}
\def\eeq{\end{equation}}
\def\bea{\begin{eqnarray}}
\def\eea{\end{eqnarray}}

\begin{document}

\title{
Non-Abelian $SU(2)$ gauge fields through 
density-wave order and strain in  graphene}

\author{Sarang Gopalakrishnan, Pouyan Ghaemi, and Shinsei Ryu}
\affiliation{Department of Physics, University of Illinois at Urbana-Champaign, 1110 West Green Street, Urbana, Illinois 61801}

\date{Apr 30, 2012}

\begin{abstract}
Spatially varying strain patterns can qualitatively alter the electronic properties of graphene, acting as effective valley-dependent magnetic fields and giving rise to pseudo-Landau-level (PLL) quantization~\cite{Novoselov05}. Here, we show that the strain-induced magnetic field is one component of an $SU(2)$ non-Abelian gauge field within the low-energy theory of graphene, and identify the other two components as period-3 charge-density waves. We show that these density-waves, if spatially varied, give rise to PLL quantization. We also argue that strain-induced magnetic fields can induce density-wave order in graphene, thus dynamically gapping out the lowest PLL; moreover, the ordering should generically be accompanied by dislocations. We discuss experimental signatures of these effects.
\end{abstract}

\maketitle

The discovery of graphene~\cite{Novoselov05}---a carbon monolayer with low-energy electronic properties governed by the Dirac equation~\cite{mpr}---has stimulated enormous interest in condensed matter systems having Dirac quasiparticles. 
Although other systems supporting Dirac quasiparticles have subsequently been discovered (e.g., the surface bands of topological insulators~\cite{qi,hasankane}), graphene is uniquely tunable through lattice deformations and strains, being a soft two-dimensional membrane~\cite{mpr}. Strains alter the electronic structure of graphene by modulating the hopping amplitudes between neighboring lattice sites. 
A striking example of the electronic consequences of strain is that certain spatially varying patterns of deformation can mimic the effects of a pseudo-``magnetic field'' that has opposite signs in the vicinity of the two Dirac points (which we shall refer to as the valleys $\textbf{K}$ and $\textbf{K}'$~\cite{guinea2010}). Such pseudo-magnetic fields, on the order of $300$ Tesla, were recently demonstrated in pioneering experiments~\cite{levy2010} on nanoscale graphene bubbles; more recently, effective fields of order $60$ Tesla were realized by artificially designing a lattice of carbon-monoxide molecules on copper~\cite{manoharan}. At these fields, each valley is deep in the quantum Hall regime, so that its electronic structure consists of well-spaced pseudo-Landau levels (PLLs).
Because the PLLs are highly degenerate, one expects correlation effects to be strong within them; indeed, recent works have shown that, in the presence of interactions, partially-filled PLLs are unstable to forming ordered states such as valley ferromagnets, spin-Hall phases and triplet superconductors~\cite{pouyanst,dimadima}.

\begin{figure}
\centering
\includegraphics{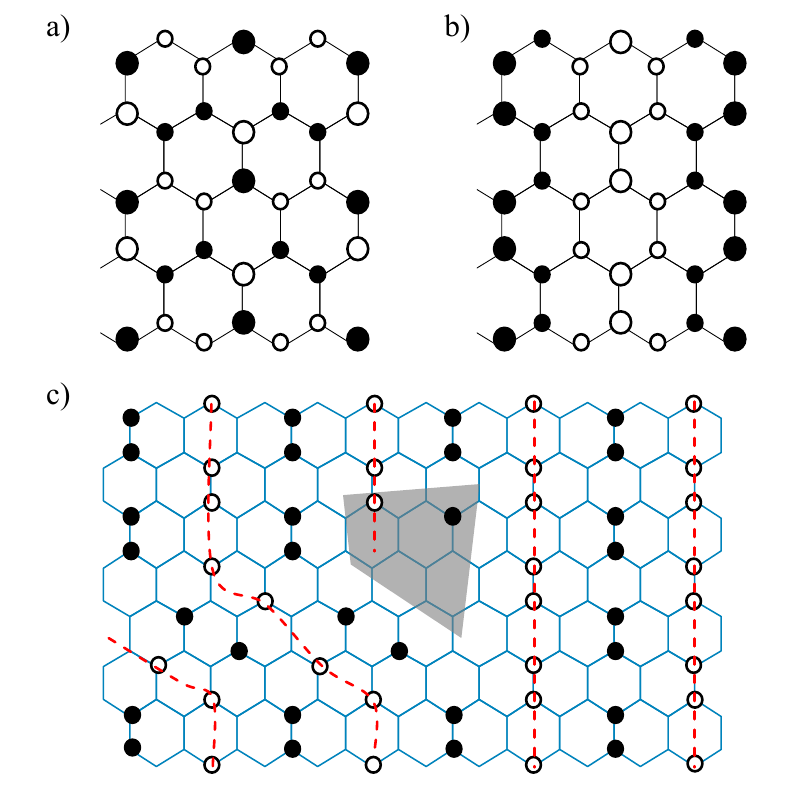}
\caption{
$SU(2)$ gauge potentials [i.e., 
charge-density-wave 
(3CDW) patterns] 
and their defects. Panels (a) and (b) show the patterns corresponding to $\zeta_{x,1}$ and $\zeta_{y,2}$ respectively (see Table~\ref{table1} for definitions); the other two patterns are related to these by translation. 
Black (white) circles denote excess positive (negative) charge; 
larger circles correspond to greater excess charge. Panel (c) sketches a dislocation in 
the 3CDW pattern; 
the dislocation core is marked in gray. 
Dashed red lines trace the crests of the 3CDW pattern.
}
\label{fig:honeycomb}
\end{figure}

\begin{figure}
	\centering
		\includegraphics[height=4in, width=3.4in]{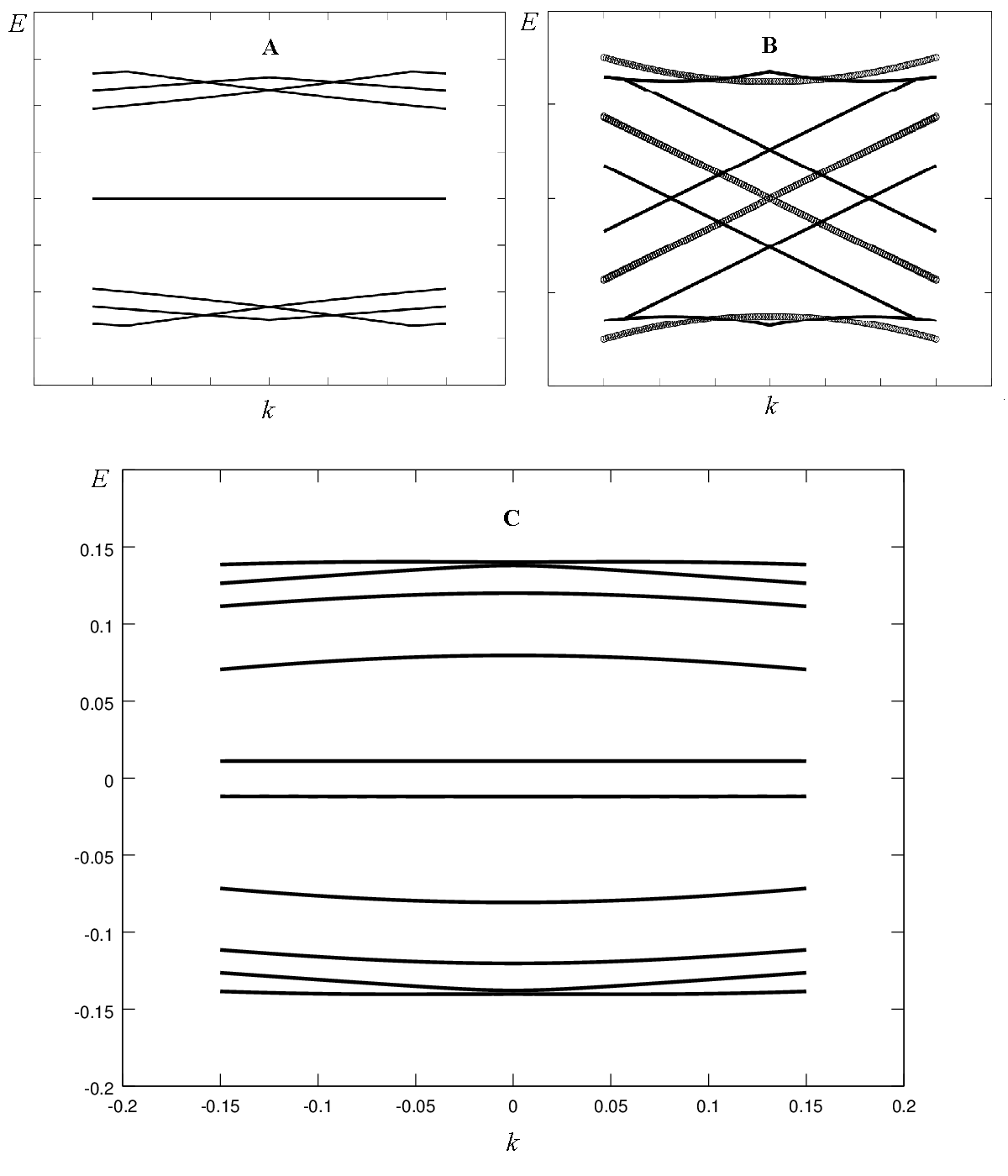}
	\caption{
(A) The band structure in the presence of strain, calculated for a 100-site thick graphene nanoribbon with zigzag edges, showing pseudo-Landau levels (PLLs). All energies are computed in terms of the unperturbed hopping matrix element; in the presence of strain, the hopping along one of the three directions increases linearly from $t$ to $1.9t$ across the sample. Note that the zero-energy level is in fact fourfold degenerate, including both two zero-energy PLLs and two modes localized on zigzag edges. 
(B) The location  of the Dirac cones in the absence of strain, both with the perturbation $\zeta_{x,i}$ (thin solid line) and without it (thick line). As noted in the main text, constant gauge potentials are expected to shift the Dirac cones. 
(C) The band structure 
in the presence of both strain and the perturbation $\zeta_{x,i}$. 
Note that the fourfold-degenerate zero-energy mode is completely split, with the two more widely separated levels corresponding to the lowest PLL. (This was checked by explicitly computing the wavefunctions.)}
	\label{fig:fig1}
\end{figure}

\begin{figure}
	\centering
		\includegraphics[height=2.8in, width=3.4in]{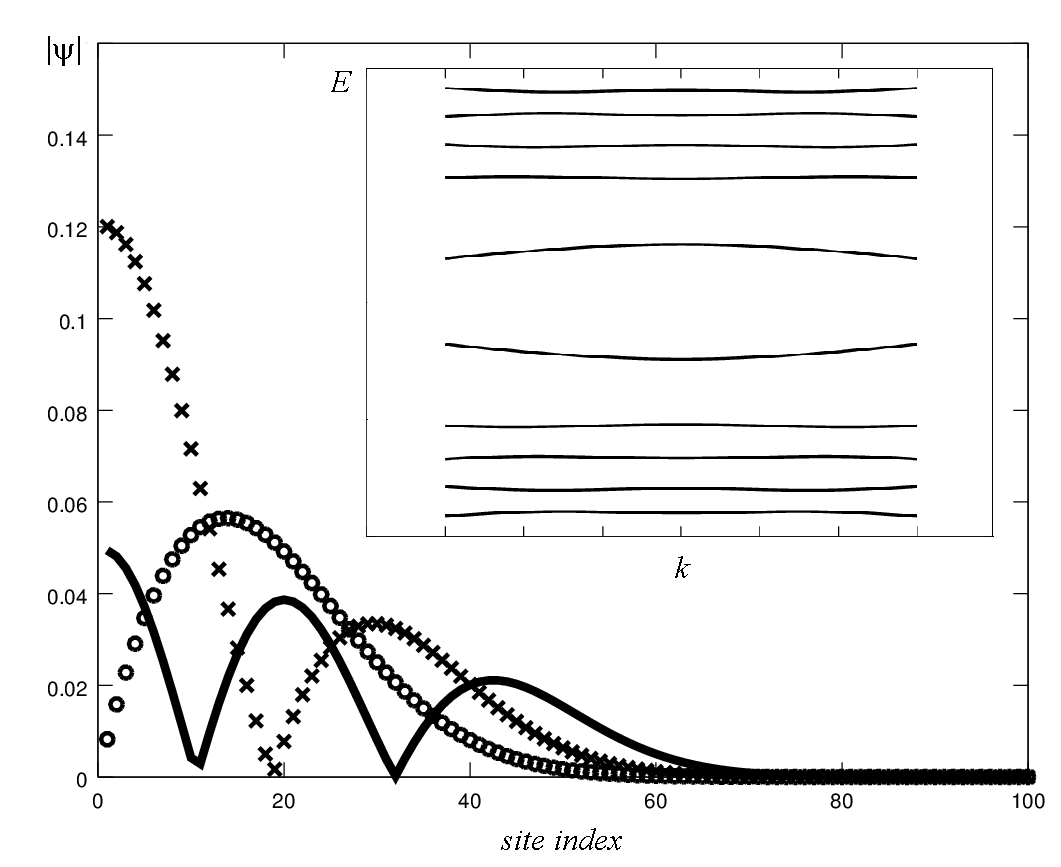}
	\caption{Pseudo-Landau levels (PLLs) generated by the $SU(2)$ pseudo-gauge potential 
$A^2_x Q_2 = B Q_2 y$ on a 100-site nanoribbon with zigzag edges. The value of $Q_2$ varies linearly from zero at one end of the ribbon to $0.74 t$ at the other end. 
The main panel shows wavefunctions in the first, second, and third distinct PLL, calculated on a 100-site nanoribbon with zigzag edges. The inset shows the overall level structure (cf. Fig.~\ref{fig:fig1}A, for the case of regular strain), demonstrating that the PLLs are indeed flat for momenta near the Dirac cone.}
	\label{fig:fig-ppll}
\end{figure}

In the present work we show that the strain-induced, valley-dependent magnetic field is one component of a 
non-Abelian $SU(2)$ gauge field 
within the low-energy theory of graphene. We identify the other two generators of this $SU(2)$ gauge field as 
period-3 charge-density waves 
(3CDWs)
(Fig.~\ref{fig:honeycomb})
that mix the $\mathbf{K}$ and $\mathbf{K}'$ valleys. We show that these charge-density waves act as gauge potentials: when their amplitude is constant, they move the Dirac cones
[Fig.~\ref{fig:fig1}(B)]; 
but when their amplitude is spatially varied, they can give rise to Landau-level quantization, as shown in Fig.~\ref{fig:fig-ppll}. Although methods for realizing non-Abelian $SU(2)$ gauge fields had previously been proposed in ultracold atomic settings~\cite{osterloh, spielman, lewenstein} and in twisted bilayer graphene~\cite{guinea:bilayer}, such fields have yet to be experimentally realized in condensed matter. The present work suggests an alternative approach, which might be easier to implement, e.g., in molecular graphene~\cite{manoharan}.

Having established the $SU(2)$ gauge structure, 
we turn to the effects of 
the 3CDW patterns on the 
PLL
structure induced by strain. 
We find that, although these perturbations 
do not open up gaps in unstrained graphene, they \textit{do }gap out 
the lowest (i.e., zero-energy) PLL. 
On general grounds, then, we expect these gaps, 
and the corresponding 
3CDW patterns, 
to be dynamically generated by electron-electron or electron-phonon interactions whenever graphene is strained, as they would reduce the ground-state energy. (The relation between different mass gaps and the corresponding ordered states was previously explored, for unstrained graphene, in Refs.~\cite{ps1,ps2,shinsei09}.)

After discussing the 
3CDW patterns, 
we turn to their defects (i.e., dislocations), and show that these defects are entwined with the ordering in a distinctive way, owing to the valley-dependence of the pseudo-magnetic field. In contrast with the case of a regular field, for a strain-induced field a uniform 3CDW perturbation does \textit{not} mix the spatially coincident Landau orbitals in the two valleys, as these are \textit{counter-propagating}. However, 
a 3CDW perturbation 
with a defect at the origin can mix the valleys and open up a gap. Thus, in experimental geometries such as that of Ref.~\cite{levy2010}, the defects as well as the order are likely to be dynamically generated.  

\textit{Model.}  In the absence of interactions, the tight-binding Hamiltonian of strained graphene reads
\begin{equation}
H_0 =\sum_{\mathbf{r}_{i}} \sum_{a=1,2,3} (t+\delta t_{a} (\mathbf{r}_{i})) ( a^\dagger (\mathbf{r}_{i}) b (\mathbf{r}_{i} + \mbox{\boldmath$\delta$}_a) + h.c. ),
\label{hamiltonian1}
\end{equation}
where $\delta t_{a} (\mathbf{r}_{i})$ is the strain-induced variation of the nearest neighbour hopping amplitude between the $A$-sublattice site at $\mathbf{r}_{i} $ and the $B$-sublattice site at $\mathbf{r}_{i} + \mbox{\boldmath$\delta$}_a$ of the bipartite honeycomb lattice \cite{mpr}. The vectors $\mbox{\boldmath$\delta$}_{a}$ connect any $A$-sublattice atom to its three $B$-sublattice nearest neighbors. In the absence of strain, the low-energy excitations correspond to linearly dispersing states close to the two Dirac points at momenta $\pm {\bf K}$ with ${\bf K}=(4 \pi/3 \sqrt{3}a_0) {\bf e_x}$, $a_0$ being the carbon-carbon bond length \cite{mpr}.
Near the Dirac points $\mathbf{K}$ and $\mathbf{K}'$ the wavefunctions of such states can be written as four-component spinors 
$\Psi \equiv (\psi_{A,\mathbf{K}}, \psi_{B,\mathbf{K}}, \psi_{A,\mathbf{K}'}, 
\psi_{B,\mathbf{K}'})$ 
where the first index denotes the component of the wavefunction on the $A(B)$ sublattice of the honeycomb unit cell, and the second index denotes the component of the state that is associated with the $\mathbf{K}$ ($\mathbf{K}'$) valley. The low energy effective Hamiltonian close to the Dirac points reads as: 
\beq
\label{kn}
\mathcal{H}_0 = v_F \left[\hat{p}_x \Gamma_x + \hat{p}_y \Gamma_y\right]
\eeq
where $\Gamma_x =\tau_3 \sigma_1$, $\Gamma_y = \tau_0 \sigma_2$, $v_F$ is the Fermi velocity, and the $\sigma$ and $\tau$ operators are Pauli matrices acting on sublattice and valley indices respectively. 
We have 
not included
the physical spin index as it does not affect our analysis, provided that the spin-orbit coupling is negligible.


There are three terms in the low-energy theory (i.e., ``charges'') 
that commute with 
the Hamiltonian (\ref{kn}): 
$ Q_1 \equiv -\tau_2 \sigma_2,
\quad Q_2 \equiv \tau_1 \sigma_2, \quad Q_3 \equiv \tau_3 \sigma_0 $. 
These realize 
an $SU(2)$ 
pseudo-spin algebra $\left[Q_i, Q_j\right] = 2 {i}\epsilon_{ijk} Q_j$. 
We also define the electromagnetic $U(1)$  charge  $Q_0 \equiv \tau_0 \sigma_0$ (i.e., the identity operator), which commutes with the other charges. We can minimally couple $\mathcal{H}_0$ to the gauge potentials associated with these charges, thus arriving at the Hamiltonian:
\beq\label{su2fields}
\mathcal{H} = v_F \left[\Gamma_x \left(
\hat{p}_x - \sum_{i = 0}^3 A_x^i Q_i\right) 
+ \Gamma_y \left(
\hat{p}_y - \sum_{i = 0}^3 A_y^i Q_i\right)\right].
\eeq
We turn to the microscopic origins of the $SU(2)$ gauge potentials, $\zeta_{\mu,i} \equiv \Gamma_\mu Q_i$, where $\mu = x,y$ and $i = 1,2,3$. Of these, $\zeta_{\mu,3}$ comprise the familiar strain-induced vector potential. Strain generates a gauge field 
given by $A_x^0+i A_y^3=\sum_{a=1,2,3} \delta t_{a}(\textbf{r})e^{\pm i \textbf{K} \cdot \mbox{\boldmath$\delta$}_{a}}$ near the Dirac points $\pm \mathbf{K}$.
 Note that $\sum_{a=1,2,3} \delta t_{a}(\textbf{r})e^{\pm i \textbf{K} \cdot \mbox{\boldmath$\delta$}_{a}}$ is complex because the nearest-neighbor hoppings are not symmetric under inversion. The real part of the strain gauge field $ A_x^0$ is the same in both valleys and therefore couples to  $Q_0$; it can be gauged away assuming time-reversal symmetry holds. On the other hand, the imaginary part $i A_y^3$ has opposite sign in the two valleys and couples to $Q_3$  leading  to the valley-dependent magnetic fields realized in the experiments of Ref.~\cite{levy2010}.


The four remaining gauge potentials (see Table \ref{table1})
originate, as we shall now see, as 3CDWs.
That they should be charge density waves 
can be seen as follows: (a)~the perturbations mix the valleys, and must therefore involve spatial modulations that enlarge the unit cell; (b)~they do not mix the sublattices (i.e., they are proportional either to $\sigma_3$ or to $\sigma_0$), and can therefore include only on-site charge offsets and intra-sublattice (e.g., next-nearest neighbor) hopping. Two simple perturbations satisfying both criteria are charge modulations of wavevector $\mathbf{G}$ (where $\mathbf{G} \equiv \mathbf{K} - \mathbf{K}'$ is a vector connecting the two Dirac points), which realize $\tau_1$ and $\tau_2$ respectively:
\beq\label{taus}
\tau_1 \leftrightarrow \cos(\mathbf{G \cdot r}), \quad \tau_2 \leftrightarrow \sin(\mathbf{G \cdot r}).
\eeq
The gauge potentials $\zeta_{x,i}$ are realized when the density waves on 
$A$ and $B$ sublattices 
are $\pi$ out of phase, whereas the potentials $\zeta_{y,i}$ are realized when the density waves 
on the $A$ and $B$ sublattices 
are in phase. Fig.~\ref{fig:honeycomb} shows the two corresponding density-wave arrangements, which are listed in Table~\ref{table1}. 

\begin{table}
  \begin{tabular}{| c || c | c | c | }
		\hline
		Term & Low-energy & LPLL & Microscopic  \\
		\hline 
		$\zeta_{x,1}$ & $\tau_1 \sigma_3$ & $\tau_1$ & $\cos(\mathbf{G \cdot r}) \cos\left[ \pi(y/a_0 + \frac{1}{4}) \right]$ \\
		$\zeta_{x,2}$ & $\tau_2 \sigma_3$ & $\tau_2$ & $\sin(\mathbf{G \cdot r}) \cos\left[ \pi(y/a_0 + \frac{1}{4}) \right] $ \\
		$\zeta_{y,1}$ & $\tau_2 \sigma_0$ & $\tau_2$ & $\sin(\mathbf{G \cdot r})$ \\
		$\zeta_{y,2}$ & $\tau_1 \sigma_0$ & $\tau_1$ & $\cos(\mathbf{G \cdot r})$ \\
		\hline
  \end{tabular}
  \caption{
Density-wave-based gauge potentials in graphene. 
This table lists the microscopic and low-energy forms of the perturbations, as well as the forms after projection onto the lowest pseudo-Landau level (LPLL).}\label{table1}
\end{table}

Numerical band structure calculations on nano-ribbons including these terms are shown 
in Figs.~\ref{fig:fig1} and \ref{fig:fig-ppll}; 
as one can see, the $\zeta$'s do not open up gaps in the absence of strain, but do shift the Dirac points in momentum space, as a vector potential is expected to do. Moreover, if the coefficient of 
$\zeta_{y,2}$, say,
is varied linearly with $x$ 
(cf. the Landau gauge description of a uniform magnetic field), it gives rise to PLL quantization as shown in Fig.~\ref{fig:fig-ppll}. 
We have thus established that
these terms can be regarded, 
together with the strain-induced gauge potentials, 
as enabling the realization of a general $SU(2)$ gauge field.


\textit{Strain-induced PLLs and their mass gaps}. 
The pseudospin $SU(2)$ symmetry (at low energies)
allowed us to treat the strain and 
3CDWs on the same footing in the preceding discussion.
We now break this symmetry by considering 
the PLL structure created by a strain pattern
inducing a uniform pseudo-magnetic field.
The eigenstates of $\mathcal{H}$ then fall into PLLs
at energies $E_n \sim \sqrt{n}$, 
and in particular there is a PLL at zero energy 
in each valley \cite{mpr}, 
which we term the LPLL.
While the LPLL shares some features with
the zero-energy Landau levels induced by a real magnetic field,
it is distinct in two essential ways, as follows.
(1)~In contrast with the case of a real magnetic field 
(in which the two lowest Landau level wavefunctions in the two valleys are located on opposite sublattices), 
the LPLL wavefunctions in both valleys 
are located entirely on the $A$ sublattice~\cite{pouyanst}. 
(2)~The wavefunctions in the LPLL
in the $\mathbf{K}$($\mathbf{K}'$) valleys
have the four-component form 
$\Psi^{\mathbf{K}}_{0,m} = (\phi_{0,m},0,0,0)$ 
and 
$\Psi^{\mathbf{K}'}_{0,m} = (0,0,\phi_{0,m}^*,0)$,
respectively, 
where $\phi_{0,m}$ is the $m$th Landau orbital in the lowest 
(nonrelativistic) Landau level
(see below). [By contrast, in the case of a real magnetic field, the second of these would be $\Psi^{\mathbf{K}'}_{0,m} = (0,0,0,\phi_{0,m})$.]
Thus, in a pseudo-magnetic field, the PLL orbitals are \textit{counter-propagating}, whereas in a real magnetic field they are \textit{co-propagating}. Consequently, for a \textit{real} magnetic field, the valley index can be regarded as a mere flavor degree of freedom, and a uniform valley-mixing perturbation mixes the Landau orbitals in the two valleys. However, for \textit{strained} graphene,
the valley index
should not be interpreted as a flavor index, because the wavefunctions in the two valleys are spatially distinct.
As we shall see, this implies
that inhomogeneous mass terms (i.e., mass terms with defects) are necessary
to gap the LPLL. 
The precise consequences depend 
on \textit{which gauge} in pseudo-vector potentials
is simulated by the strain pattern.
(This is possible because, while two gauge-equivalent electromagnetic vector potentials are physically identical, 
different pseudo-vector potentials are physically distinct
as they correspond to different patterns of strain.)


\textit{Uniform perturbations.}
We now address point~(1) from the previous paragraph, ignoring spatial structure and considering the effects of uniform (i.e., defect-free) 3CDW perturbations precisely \textit{at} zero momentum (i.e., \textit{at} the Dirac point).
%
Many of the properties of the LPLL follow from the triviality of its sublattice structure: in particular, one can find the form of any perturbation in the LPLL by projecting it onto the $A$ sublattice. Thus, perturbations within the LPLL are completely described by $2\times 2$ matrices in $\tau$ (i.e., valley) space 
(Table~\ref{table1}).
As a consequence, several perturbations that open up gaps in \textit{unstrained} graphene, such as an intra-unit-cell 
charge-density wave
($\sigma_3 \tau_0$) and the Kekul\'{e} distortion ($\sigma_1 \tau_1$), are trivial when projected to the LPLL. 
However, the $SU(2)$ gauge potentials $\zeta_{\mu,i}$ \textit{do} open up gaps within the LPLL, 
as they project onto either $\tau_1$ or $\tau_2$ 
(Table~\ref{table1}), 
and can thus mix LPLL orbitals from the $\mathbf{K}$ and $\mathbf{K}'$ valleys.

The two further perturbations that split 
the LPLL are valley polarization, $m_p \equiv \tau_3 \sigma_0$, and the Haldane mass~\cite{haldane}, $m_H \equiv \tau_3 \sigma_3$. Within the LPLL, these terms are \textit{equivalent}; both correspond to $\tau_3$, which shifts the energy of one valley with respect to the other. 
Both terms anticommute with the gauge fields; therefore, within the low-energy theory, it seems that these masses can be continuously deformed into one another. (Thus, the fate of the topologically-protected edge mode associated with $m_H$ cannot be addressed within the low-energy theory. Numerical calculations of the band structure in the presence of both $m_H$ and 
$\zeta_{x,i}$ suggest that as $\zeta_{x,i}$
is increased, the edge state drifts away from the Dirac points; it is therefore plausible that 
$\zeta_{x,i}$ and $m_H$ compete away from the Dirac points. We shall revisit this question in future work.)

The perturbations discussed above are likely to be dynamically generated in experiments with neutral graphene (i.e., a half-filled LPLL), as opening up a gap would lower the ground-state energy. 
The 3CDWs
could arise either because of the electron-phonon coupling or the electron-electron coupling; moreover, a spontaneous valley polarization might arise due to electron-electron interactions~\cite{pouyanst}. Alternatively, one can externally impose these gaps by growing graphene on an appropriately patterned substrate. 

\textit{Momentum dependence}. We now turn to the second distinctive feature of the PLLs [point~(2) above] and discuss the role played by the spatial structure of PLL wavefunctions.  
In order to address this, we shall consider the nature of the 
3CDW perturbations on
orbitals \textit{away} from the Dirac point.
%

We first consider the strain pattern realizing the Landau gauge;
here, 
if the 3CDW amplitude is uniform,
the left-valley Landau orbital 
$\phi_{0,m}(x,y) \propto 
e^{i \frac{2\pi m}{L} x}e^{-\left(y-\frac{2\pi m}{L}l^2\right)^2/l^2}$
($l$ being the magnetic length) 
can hybridize \textit{only} with the right-valley Landau orbital 
$\phi^*_{0,-m}(x,y) \propto 
e^{i \frac{2\pi m}{L} x}e^{-\left(y+\frac{2\pi m}{L}l^2\right)^2/l^2}$,
due to the conservation of $k_x$.
Although the overlap between these two states is nonzero, it \textit{decreases} as $|m|$ is increased, and becomes exponentially small for $m l \gg L$. This decrease of overlap leads to the convexity of the LPLL gap shown in Fig.~\ref{fig:fig1}, and implies that the 
3CDW 
gap is an inherently mesoscopic phenomenon. (However, note that all experimental realizations of pseudo-magnetic fields in strained graphene involve mesoscopic systems.) 

\textit{Point defects}.
If the strain realizes a symmetric gauge pattern, 
 $\phi_{0,m}(r,\theta) \propto e^{i m\theta}r^m e^{-r^2/l^2}$,
as in the experiments of Ref.~\cite{levy2010}, 
the consequences are even more striking.  
Here, the only allowed orbitals have $m \geq 0$,
because
otherwise $\phi_{0,m}$ 
is not normalizable. As a result, \textit{no }uniform perturbation can gap out the $m \neq 0$ orbitals. Thus, in order to open up a gap, it is necessary to consider configurations in which the perturbations have defects, i.e., edge dislocations 
in the 3CDW case.
Within the LPLL, 
the two independent 
3CDW orders 
are represented by $\tau_1$ and $\tau_2$ 
(Table~\ref{table1}). 
The 3CDW
can support edge dislocations 
[Fig.~\ref{fig:honeycomb}(c)],
 which take the form of vortex solutions within the LPLL theory:
\beq
\Delta(r) 
\left[
\tau_1 \cos(n\theta)+
\tau_2 \sin(n\theta)\right]
= \left( \begin{array}{cc} 0 & \Delta(r) e^{in\theta} \\ \Delta(r) e^{-in\theta} & 0 \end{array} \right) 
\eeq
where $\Delta(r)$ is a function that vanishes at $r = 0$ and is constant for $r \gg l_M$. 
Such vortex solutions carry angular momentum, 
so that
they can lead to hybridization of different orbitals in the LPLL:
\begin{align}
&
\int d\theta dr 
\left(\phi^*_{0,m},0 \right) \Delta(r) 
\left[
\tau_1 \cos(n\theta)+\tau_2 \sin(n\theta)
\right]
\left(\begin{array}{c} 0 \\  
\phi^*_{0,m'} \end{array}\right) 
\nonumber \\
&\quad  \propto \delta_{0,n-m-m'}
\end{align}
for vorticity $n$, the Landau orbitals with $m<n$ will be gapped. As a result, a larger vorticity can gap out more Landau orbitals and will thus lead to lower energy (although a larger vorticity might also cost greater electrostatic or elastic energy). 
In this case, one expects the defects to be dynamically generated 
along with the 3CDW patterns.

\textit{Effects in higher PLLs}. 
We now turn to the effects of the aforementioned perturbations when projected onto higher PLLs. In the higher PLLs, the sublattice structure is not trivial, 
so that the six masses are \textit{distinct} perturbations. 
These perturbations fall into two classes, depending on whether their sublattice component is $\sigma_0$ or $\sigma_3$. Perturbations of the former class split all the PLLs. For the valley polarization term, this is obvious by inspection: a polarization of the form $\tau_z$ moves all the PLLs in the $\mathbf{K}$ up and all those at $\mathbf{K}'$ down, thus opening up a gap of size $2\alpha$ between any pair of PLLs. Similarly, degenerate perturbation theory shows that the perturbations $\zeta_{y,i}$ mix 
$\Psi^{\mathbf{K}}_{n,m}$ and $\Psi^{\mathbf{K}'}_{n,m}$ (assuming, for simplicity, that the strain pattern realizes the Landau gauge). 
By contrast, $m_H$ and $\zeta_{x,i}$ do \textit{not} mix 
$\Psi^{\mathbf{K}}_{n,m}$ and $\Psi^{\mathbf{K}'}_{n,m}$, 
and thus preserve the double degeneracy of higher PLLs. 
However, these perturbations \textit{do} mix 
$\Psi^{\mathbf{K}}_{n,m}$ and $\Psi^{\mathbf{K}'}_{-n,m}$, 
thus shifting the energy of the $n$th PLL by $\alpha^2 / (2 \Delta_0 \sqrt{n})$, where $\Delta_0$ is the cyclotron energy scale. 

These considerations might influence which set of perturbations is in fact dynamically generated. In particular, if one of the nonzero PLLs is half-filled, the favored perturbations are those of the $\sigma_0$ class. By contrast, at half-filling of the LPLL (i.e., for neutral graphene), the above argument suggests that the $\sigma_3$ class of perturbations is preferred, as these lower the energy of the filled negative-energy PLLs. It is not clear, however, whether this energy saving is outweighed by changes in band structure away from the Dirac points. 

\textit{Experimental aspects}. We close by touching upon various experimental considerations.
The 
 3CDW ordering described here can either be realized spontaneously in the presence of strain, or imposed externally. In the former case, the pattern of ordering can be detected easily via scanning-tunneling microscopy (STM); this technique is commonly used to study stripe ordering, e.g., in high-temperature superconductors (see, e.g., Ref.~\cite{kapitulnik}). 
Alternatively, one can study 
the 3CDW 
perturbations by \textit{externally imposing} them. 
This is easiest to do in the case of engineered systems, such as molecular graphene~\cite{manoharan}, where a 
3CDW pattern such as $\zeta_{y,i}$ can be imposed by hand on the triangular network of adsorbate molecules (or, alternatively, for hexagonal optical lattices~\cite{sengstock}). 
However, it might also be possible to realize it using systems of graphene grown on anisotropic substrates, which favor 
3CDW formation. 


\textit{Acknowledgments}. 
S.G. and P.G. are indebted to Paul Goldbart for helpful discussions. The authors acknowledge support from DOE DE-FG02-07ER46453 (S.G.) and ICMT at UIUC (P.G.).


\begin{thebibliography}{19}%
\makeatletter
\providecommand \@ifxundefined [1]{%
 \@ifx{#1\undefined}
}%
\providecommand \@ifnum [1]{%
 \ifnum #1\expandafter \@firstoftwo
 \else \expandafter \@secondoftwo
 \fi
}%
\providecommand \@ifx [1]{%
 \ifx #1\expandafter \@firstoftwo
 \else \expandafter \@secondoftwo
 \fi
}%
\providecommand \natexlab [1]{#1}%
\providecommand \enquote  [1]{``#1''}%
\providecommand \bibnamefont  [1]{#1}%
\providecommand \bibfnamefont [1]{#1}%
\providecommand \citenamefont [1]{#1}%
\providecommand \href@noop [0]{\@secondoftwo}%
\providecommand \href [0]{\begingroup \@sanitize@url \@href}%
\providecommand \@href[1]{\@@startlink{#1}\@@href}%
\providecommand \@@href[1]{\endgroup#1\@@endlink}%
\providecommand \@sanitize@url [0]{\catcode `\\12\catcode `\$12\catcode
  `\&12\catcode `\#12\catcode `\^12\catcode `\_12\catcode `\%12\relax}%
\providecommand \@@startlink[1]{}%
\providecommand \@@endlink[0]{}%
\providecommand \url  [0]{\begingroup\@sanitize@url \@url }%
\providecommand \@url [1]{\endgroup\@href {#1}{\urlprefix }}%
\providecommand \urlprefix  [0]{URL }%
\providecommand \Eprint [0]{\href }%
\@ifxundefined \urlstyle {%
  \providecommand \doi  [0]{\begingroup \@sanitize@url \@doi}%
  \providecommand \@doi [1]{\endgroup \@@startlink {\doibase
  #1}doi:\discretionary {}{}{}#1\@@endlink }%
}{%
  \providecommand \doi  [0]{doi:\discretionary{}{}{}\begingroup
  \urlstyle{rm}\Url }%
}%
\providecommand \doibase [0]{http://dx.doi.org/}%
\providecommand \Doi [0]{\begingroup \@sanitize@url \@Doi }%
\providecommand \@Doi  [1]{\endgroup\@@startlink{\doibase#1}\@@Doi}%
\providecommand \@@Doi [1]{#1\@@endlink}%
\providecommand \selectlanguage [0]{\@gobble}%
\providecommand \bibinfo  [0]{\@secondoftwo}%
\providecommand \bibfield  [0]{\@secondoftwo}%
\providecommand \translation [1]{[#1]}%
\providecommand \BibitemOpen [0]{}%
\providecommand \bibitemStop [0]{}%
\providecommand \bibitemNoStop [0]{.\EOS\space}%
\providecommand \EOS [0]{\spacefactor3000\relax}%
\providecommand \BibitemShut  [1]{\csname bibitem#1\endcsname}%
\bibitem [{\citenamefont {Novoselov}\ \emph {et~al.}(2005)\citenamefont
  {Novoselov}, \citenamefont {Geim}, \citenamefont {Morozov}, \citenamefont
  {Jiang}, \citenamefont {Katsnelson}, \citenamefont {Grigorieva},
  \citenamefont {Dubonos},\ and\ \citenamefont {Firsov}}]{Novoselov05}%
  \BibitemOpen
  \bibfield  {author} {\bibinfo {author} {\bibfnamefont {K.~S.}\ \bibnamefont
  {Novoselov}}, \bibinfo {author} {\bibfnamefont {A.~K.}\ \bibnamefont {Geim}},
  \bibinfo {author} {\bibfnamefont {S.~V.}\ \bibnamefont {Morozov}}, \bibinfo
  {author} {\bibfnamefont {D.}~\bibnamefont {Jiang}}, \bibinfo {author}
  {\bibfnamefont {M.~I.}\ \bibnamefont {Katsnelson}}, \bibinfo {author}
  {\bibfnamefont {I.~V.}\ \bibnamefont {Grigorieva}}, \bibinfo {author}
  {\bibfnamefont {S.~V.}\ \bibnamefont {Dubonos}}, \ and\ \bibinfo {author}
  {\bibfnamefont {A.~A.}\ \bibnamefont {Firsov}},\ }\href@noop {} {\bibfield
  {journal} {\bibinfo  {journal} {Nature},\ }\textbf {\bibinfo {volume}
  {438}},\ \bibinfo {pages} {197} (\bibinfo {year} {2005})}\BibitemShut
  {NoStop}%
\bibitem [{\citenamefont {Castro~Neto}\ \emph {et~al.}(2009)\citenamefont
  {Castro~Neto}, \citenamefont {Guinea}, \citenamefont {Peres}, \citenamefont
  {Novoselov},\ and\ \citenamefont {Geim}}]{mpr}%
  \BibitemOpen
  \bibfield  {author} {\bibinfo {author} {\bibfnamefont {A.~H.}\ \bibnamefont
  {Castro~Neto}}, \bibinfo {author} {\bibfnamefont {F.}~\bibnamefont {Guinea}},
  \bibinfo {author} {\bibfnamefont {N.~M.~R.}\ \bibnamefont {Peres}}, \bibinfo
  {author} {\bibfnamefont {K.~S.}\ \bibnamefont {Novoselov}}, \ and\ \bibinfo
  {author} {\bibfnamefont {A.~K.}\ \bibnamefont {Geim}},\ }\href@noop {}
  {\bibfield  {journal} {\bibinfo  {journal} {Rev. Mod. Phys.},\ }\textbf
  {\bibinfo {volume} {81}},\ \bibinfo {pages} {109} (\bibinfo {year}
  {2009})}\BibitemShut {NoStop}%
\bibitem [{\citenamefont {Qi}\ and\ \citenamefont {Zhang}(2011)}]{qi}%
  \BibitemOpen
  \bibfield  {author} {\bibinfo {author} {\bibfnamefont {X.-L.}\ \bibnamefont
  {Qi}}\ and\ \bibinfo {author} {\bibfnamefont {S.-C.}\ \bibnamefont {Zhang}},\
  }\href@noop {} {\bibfield  {journal} {\bibinfo  {journal} {Rev. Mod. Phys.}}
  (\bibinfo {year} {2011})}\BibitemShut {NoStop}%
\bibitem [{\citenamefont {Hasan}\ and\ \citenamefont {Kane}(2010)}]{hasankane}%
  \BibitemOpen
  \bibfield  {author} {\bibinfo {author} {\bibfnamefont {M.~Z.}\ \bibnamefont
  {Hasan}}\ and\ \bibinfo {author} {\bibfnamefont {C.~L.}\ \bibnamefont
  {Kane}},\ }\href@noop {} {\bibfield  {journal} {\bibinfo  {journal} {Rev.
  Mod. Phys.},\ }\textbf {\bibinfo {volume} {82}},\ \bibinfo {pages} {3045}
  (\bibinfo {year} {2010})}\BibitemShut {NoStop}%
\bibitem [{\citenamefont {Guinea}\ \emph {et~al.}(2010)\citenamefont {Guinea},
  \citenamefont {Katsnelson},\ and\ \citenamefont {Geim}}]{guinea2010}%
  \BibitemOpen
  \bibfield  {author} {\bibinfo {author} {\bibfnamefont {F.}~\bibnamefont
  {Guinea}}, \bibinfo {author} {\bibfnamefont {M.}~\bibnamefont {Katsnelson}},
  \ and\ \bibinfo {author} {\bibfnamefont {A.}~\bibnamefont {Geim}},\
  }\href@noop {} {\bibfield  {journal} {\bibinfo  {journal} {Nat. Phys.},\
  }\textbf {\bibinfo {volume} {6}},\ \bibinfo {pages} {30} (\bibinfo {year}
  {2010})}\BibitemShut {NoStop}%
\bibitem [{\citenamefont {Levy}\ and\ \citenamefont {al.}(2010)}]{levy2010}%
  \BibitemOpen
  \bibfield  {author} {\bibinfo {author} {\bibfnamefont {L.}~\bibnamefont
  {Levy}}\ and\ \bibinfo {author} {\bibnamefont {al.}},\ }\href@noop {}
  {\bibfield  {journal} {\bibinfo  {journal} {Science},\ }\textbf {\bibinfo
  {volume} {329}},\ \bibinfo {pages} {544} (\bibinfo {year}
  {2010})}\BibitemShut {NoStop}%
\bibitem [{\citenamefont {Gomes}\ \emph {et~al.}(2012)\citenamefont {Gomes},
  \citenamefont {Mar}, \citenamefont {Ko}, \citenamefont {Guinea},\ and\
  \citenamefont {Manoharan}}]{manoharan}%
  \BibitemOpen
  \bibfield  {author} {\bibinfo {author} {\bibfnamefont {K.~K.}\ \bibnamefont
  {Gomes}}, \bibinfo {author} {\bibfnamefont {W.}~\bibnamefont {Mar}}, \bibinfo
  {author} {\bibfnamefont {W.}~\bibnamefont {Ko}}, \bibinfo {author}
  {\bibfnamefont {F.}~\bibnamefont {Guinea}}, \ and\ \bibinfo {author}
  {\bibfnamefont {H.~C.}\ \bibnamefont {Manoharan}},\ }\href@noop {} {\bibfield
   {journal} {\bibinfo  {journal} {Nature},\ }\textbf {\bibinfo {volume}
  {483}},\ \bibinfo {pages} {306} (\bibinfo {year} {2012})}\BibitemShut
  {NoStop}%
\bibitem [{\citenamefont {Ghaemi}\ \emph {et~al.}()\citenamefont {Ghaemi} \emph
  {et~al.}}]{pouyanst}%
  \BibitemOpen
  \bibfield  {author} {\bibinfo {author} {\bibfnamefont {P.}~\bibnamefont
  {Ghaemi}} \emph {et~al.},\ }\href@noop {} {}\bibinfo {note}
  {{a}rXiv:1111.3640}\BibitemShut {NoStop}%
\bibitem [{\citenamefont {Pesin}\ and\ \citenamefont {Abanin}()}]{dimadima}%
  \BibitemOpen
  \bibfield  {author} {\bibinfo {author} {\bibfnamefont {D.}~\bibnamefont
  {Pesin}}\ and\ \bibinfo {author} {\bibfnamefont {D.}~\bibnamefont {Abanin}},\
  }\href@noop {} {}\bibinfo {note} {{a}rXiv:1112.6420}\BibitemShut {NoStop}%
\bibitem [{\citenamefont {Osterloh}\ \emph {et~al.}(2005)\citenamefont
  {Osterloh}, \citenamefont {Baig}, \citenamefont {Santos}, \citenamefont
  {Zoller},\ and\ \citenamefont {Lewenstein}}]{osterloh}%
  \BibitemOpen
  \bibfield  {author} {\bibinfo {author} {\bibfnamefont {K.}~\bibnamefont
  {Osterloh}}, \bibinfo {author} {\bibfnamefont {M.}~\bibnamefont {Baig}},
  \bibinfo {author} {\bibfnamefont {L.}~\bibnamefont {Santos}}, \bibinfo
  {author} {\bibfnamefont {P.}~\bibnamefont {Zoller}}, \ and\ \bibinfo {author}
  {\bibfnamefont {M.}~\bibnamefont {Lewenstein}},\ }\Doi
  {10.1103/PhysRevLett.95.010403} {\bibfield  {journal} {\bibinfo  {journal}
  {Phys. Rev. Lett.},\ }\textbf {\bibinfo {volume} {95}},\ \bibinfo {pages}
  {010403} (\bibinfo {year} {2005})}\BibitemShut {NoStop}%
\bibitem [{\citenamefont {Lin}\ \emph {et~al.}(2011)\citenamefont {Lin},
  \citenamefont {Jim\'{e}nez-Garc\'{i}a},\ and\ \citenamefont
  {Spielman}}]{spielman}%
  \BibitemOpen
  \bibfield  {author} {\bibinfo {author} {\bibfnamefont {Y.-J.}\ \bibnamefont
  {Lin}}, \bibinfo {author} {\bibfnamefont {K.}~\bibnamefont
  {Jim\'{e}nez-Garc\'{i}a}}, \ and\ \bibinfo {author} {\bibfnamefont
  {I.}~\bibnamefont {Spielman}},\ }\href@noop {} {\bibfield  {journal}
  {\bibinfo  {journal} {Nature},\ }\textbf {\bibinfo {volume} {471}},\ \bibinfo
  {pages} {83} (\bibinfo {year} {2011})}\BibitemShut {NoStop}%
\bibitem [{\citenamefont {Goldman}\ \emph {et~al.}(2009)\citenamefont
  {Goldman}, \citenamefont {Kubasiak}, \citenamefont {Gaspard},\ and\
  \citenamefont {Lewenstein}}]{lewenstein}%
  \BibitemOpen
  \bibfield  {author} {\bibinfo {author} {\bibfnamefont {N.}~\bibnamefont
  {Goldman}}, \bibinfo {author} {\bibfnamefont {A.}~\bibnamefont {Kubasiak}},
  \bibinfo {author} {\bibfnamefont {P.}~\bibnamefont {Gaspard}}, \ and\
  \bibinfo {author} {\bibfnamefont {M.}~\bibnamefont {Lewenstein}},\ }\Doi
  {10.1103/PhysRevA.79.023624} {\bibfield  {journal} {\bibinfo  {journal}
  {Phys. Rev. A},\ }\textbf {\bibinfo {volume} {79}},\ \bibinfo {pages}
  {023624} (\bibinfo {year} {2009})}\BibitemShut {NoStop}%
\bibitem [{\citenamefont {San-Jose}\ \emph {et~al.}()\citenamefont {San-Jose},
  \citenamefont {Gonz\'{a}lez},\ and\ \citenamefont {Guinea}}]{guinea:bilayer}%
  \BibitemOpen
  \bibfield  {author} {\bibinfo {author} {\bibfnamefont {P.}~\bibnamefont
  {San-Jose}}, \bibinfo {author} {\bibfnamefont {J.}~\bibnamefont
  {Gonz\'{a}lez}}, \ and\ \bibinfo {author} {\bibfnamefont {F.}~\bibnamefont
  {Guinea}},\ }\href@noop {} {}\bibinfo {note} {ArXiv:1110.2883
  (2011)}\BibitemShut {NoStop}%
\bibitem [{\citenamefont {Ghaemi}\ \emph {et~al.}(2010)\citenamefont {Ghaemi},
  \citenamefont {Ryu},\ and\ \citenamefont {Lee}}]{ps1}%
  \BibitemOpen
  \bibfield  {author} {\bibinfo {author} {\bibfnamefont {P.}~\bibnamefont
  {Ghaemi}}, \bibinfo {author} {\bibfnamefont {S.}~\bibnamefont {Ryu}}, \ and\
  \bibinfo {author} {\bibfnamefont {D.-H.}\ \bibnamefont {Lee}},\ }\Doi
  {10.1103/PhysRevB.81.081403} {\bibfield  {journal} {\bibinfo  {journal}
  {Phys. Rev. B},\ }\textbf {\bibinfo {volume} {81}},\ \bibinfo {pages}
  {081403} (\bibinfo {year} {2010})}\BibitemShut {NoStop}%
\bibitem [{\citenamefont {Ghaemi}\ and\ \citenamefont {Ryu}(2012)}]{ps2}%
  \BibitemOpen
  \bibfield  {author} {\bibinfo {author} {\bibfnamefont {P.}~\bibnamefont
  {Ghaemi}}\ and\ \bibinfo {author} {\bibfnamefont {S.}~\bibnamefont {Ryu}},\
  }\href@noop {} {\bibfield  {journal} {\bibinfo  {journal} {Phys. Rev. B},\
  }\textbf {\bibinfo {volume} {85}},\ \bibinfo {pages} {075111} (\bibinfo
  {year} {2012})}\BibitemShut {NoStop}%
\bibitem [{\citenamefont {Ryu}\ \emph {et~al.}(2009)\citenamefont {Ryu},
  \citenamefont {Mudry}, \citenamefont {Hou},\ and\ \citenamefont
  {Chamon}}]{shinsei09}%
  \BibitemOpen
  \bibfield  {author} {\bibinfo {author} {\bibfnamefont {S.}~\bibnamefont
  {Ryu}}, \bibinfo {author} {\bibfnamefont {C.}~\bibnamefont {Mudry}}, \bibinfo
  {author} {\bibfnamefont {C.-Y.}\ \bibnamefont {Hou}}, \ and\ \bibinfo
  {author} {\bibfnamefont {C.}~\bibnamefont {Chamon}},\ }\Doi
  {10.1103/PhysRevB.80.205319} {\bibfield  {journal} {\bibinfo  {journal}
  {Phys. Rev. B},\ }\textbf {\bibinfo {volume} {80}},\ \bibinfo {pages}
  {205319} (\bibinfo {year} {2009})}\BibitemShut {NoStop}%
\bibitem [{\citenamefont {Haldane}(1988)}]{haldane}%
  \BibitemOpen
  \bibfield  {author} {\bibinfo {author} {\bibfnamefont {F.}~\bibnamefont
  {Haldane}},\ }\href@noop {} {\bibfield  {journal} {\bibinfo  {journal} {Phys.
  Rev. Lett.},\ }\textbf {\bibinfo {volume} {61}},\ \bibinfo {pages} {2015}
  (\bibinfo {year} {1988})}\BibitemShut {NoStop}%
\bibitem [{\citenamefont {Howald}\ \emph {et~al.}(2003)\citenamefont {Howald},
  \citenamefont {Eisaki}, \citenamefont {Kaneko},\ and\ \citenamefont
  {Kapitulnik}}]{kapitulnik}%
  \BibitemOpen
  \bibfield  {author} {\bibinfo {author} {\bibfnamefont {C.}~\bibnamefont
  {Howald}}, \bibinfo {author} {\bibfnamefont {H.}~\bibnamefont {Eisaki}},
  \bibinfo {author} {\bibfnamefont {N.}~\bibnamefont {Kaneko}}, \ and\ \bibinfo
  {author} {\bibfnamefont {A.}~\bibnamefont {Kapitulnik}},\ }\href@noop {}
  {\bibfield  {journal} {\bibinfo  {journal} {Proc. Nat. Acad. Sci},\ }\textbf
  {\bibinfo {volume} {100}},\ \bibinfo {pages} {9705} (\bibinfo {year}
  {2003})}\BibitemShut {NoStop}%
\bibitem [{\citenamefont {Soltan-Panahi}\ \emph {et~al.}(2011)\citenamefont
  {Soltan-Panahi}, \citenamefont {Struck}, \citenamefont {Hauke}, \citenamefont
  {Bick}, \citenamefont {Plenkers}, \citenamefont {Meineke}, \citenamefont
  {Becker}, \citenamefont {Windpassinger}, \citenamefont {Lewenstein},\ and\
  \citenamefont {Sengstock}}]{sengstock}%
  \BibitemOpen
  \bibfield  {author} {\bibinfo {author} {\bibfnamefont {P.}~\bibnamefont
  {Soltan-Panahi}}, \bibinfo {author} {\bibfnamefont {J.}~\bibnamefont
  {Struck}}, \bibinfo {author} {\bibfnamefont {P.}~\bibnamefont {Hauke}},
  \bibinfo {author} {\bibfnamefont {A.}~\bibnamefont {Bick}}, \bibinfo {author}
  {\bibfnamefont {W.}~\bibnamefont {Plenkers}}, \bibinfo {author}
  {\bibfnamefont {G.}~\bibnamefont {Meineke}}, \bibinfo {author} {\bibfnamefont
  {C.}~\bibnamefont {Becker}}, \bibinfo {author} {\bibfnamefont
  {P.}~\bibnamefont {Windpassinger}}, \bibinfo {author} {\bibfnamefont
  {M.}~\bibnamefont {Lewenstein}}, \ and\ \bibinfo {author} {\bibfnamefont
  {K.}~\bibnamefont {Sengstock}},\ }\href@noop {} {\bibfield  {journal}
  {\bibinfo  {journal} {Nat. Phys.},\ }\textbf {\bibinfo {volume} {7}},\
  \bibinfo {pages} {434} (\bibinfo {year} {2011})}\BibitemShut {NoStop}%
\end{thebibliography}

%

\end{document}